\documentclass[a4paper]{spie}  
\usepackage{bm}
\usepackage{multicol}
\usepackage{sidecap}
\usepackage{etoolbox}
\usepackage{tikz}
\usetikzlibrary{calc,angles}

\newcommand{\degree}[0]{^{\circ}}
\definecolor{emerald}{rgb}{0.1, 0.75, 0.55}
\definecolor{purple spa}{rgb}{0.8, 0.1, 0.6}
\definecolor{pink}{rgb}{1, 0.5, 0.5}

\newcommand{\GHcolor}[1]{\textcolor{cyan}{#1}}
\newcommand{\IFcolor}[1]{\textcolor{orange}{#1}}

\usepackage{amsmath,amsfonts,amssymb}
\usepackage{graphicx}
\usepackage[colorlinks=true, allcolors=blue]{hyperref}

\begin{document} 

\title{ Polarization effects on high contrast imaging: measurements on THD2 Bench}

\author[a]{Pierre Baudoz }
\author[b,c]{Celia Desgrange}
\author[a]{Rapha\"el Galicher}
\author[a]{Iva Laginja}
\affil[a]{LESIA, Observatoire de Paris, Universit\'e PSL, CNRS, Sorbonne Universit\'e, Universit\'e Paris Cit\'e, 5 place Jules Janssen, 92195 Meudon, France}
\affil[b]{Universit\'e Grenoble Alpes, CNRS, IPAG, F-38000 Grenoble, France}
\affil[c]{Max Planck Institute for Astronomy, K\"onigstuhl 17, D-69117 Heidelberg, Germany}

\authorinfo{Further author information: (Send correspondence to pierre.baudoz@obspm.fr)}

\pagestyle{empty} 
\setcounter{page}{301} 
 
\maketitle

\begin{abstract}
The spectroscopic study of mature giant planets and low mass planets (Neptune-like, Earth-like) requires instruments capable of achieving very high contrasts ($10^{-10}-10^{-11}$) at short angular separations. To achieve such high performance on a real instrument, many limitations must be overcome: complex component defects (coronagraph, deformable mirror), optical aberrations and scattering, mechanical vibrations and drifts, polarization effects, etc.

To study the overall impact on a complete system representative of high contrast instruments, we have developed a test bench at Paris Observatory, called THD2. In this paper, we focus on the polarization effects that are present on the bench which creates differential aberrations between the two linear polarization states. We compare the recorded beam positions of the two polarization states with the predicted from the Goos-H\"anchen and Imbert-Fedorov effects, both of which cause spatial shifts and angular deviations of the beam, longitudinal and transverse respectively. Although these effects have already been studied in the literature from the optical and quantum mechanical points of view, their measurement and impact on a complete optical bench are rather rare, although they are crucial for high-contrast instruments.

After describing the Goos-H\"anchen and Imbert-Fedorov effects and estimating their amplitude on the THD2 bench, we present the protocol we used to measure these effects of polarization on the light beam. We compare predictions and measurements and we conclude on the most limiting elements on our bench polarization-wise. 
\end{abstract}


\section{CONTEXT: Polarization effects in a high-contrast instrument}
\label{sec:intro}  
 
Reflection of vector electromagnetic wave (light) from dielectric and metal reflection surfaces causes both polarized amplitude and phase changes across wavefront, which affects image quality\cite{Breckinridge2015}. Since high contrast imaging instruments are highly sensitive to aberrations, these polarization effects have large impact on the instruments aiming to directly detect of exoplanets\cite{Breckinridge2004, Breckinridge2020}. In this paper, we study the polarization limitation of our high contrast imaging bench called THD2. 

\begin{figure}[h]
\centering
    \includegraphics[width=1\linewidth]{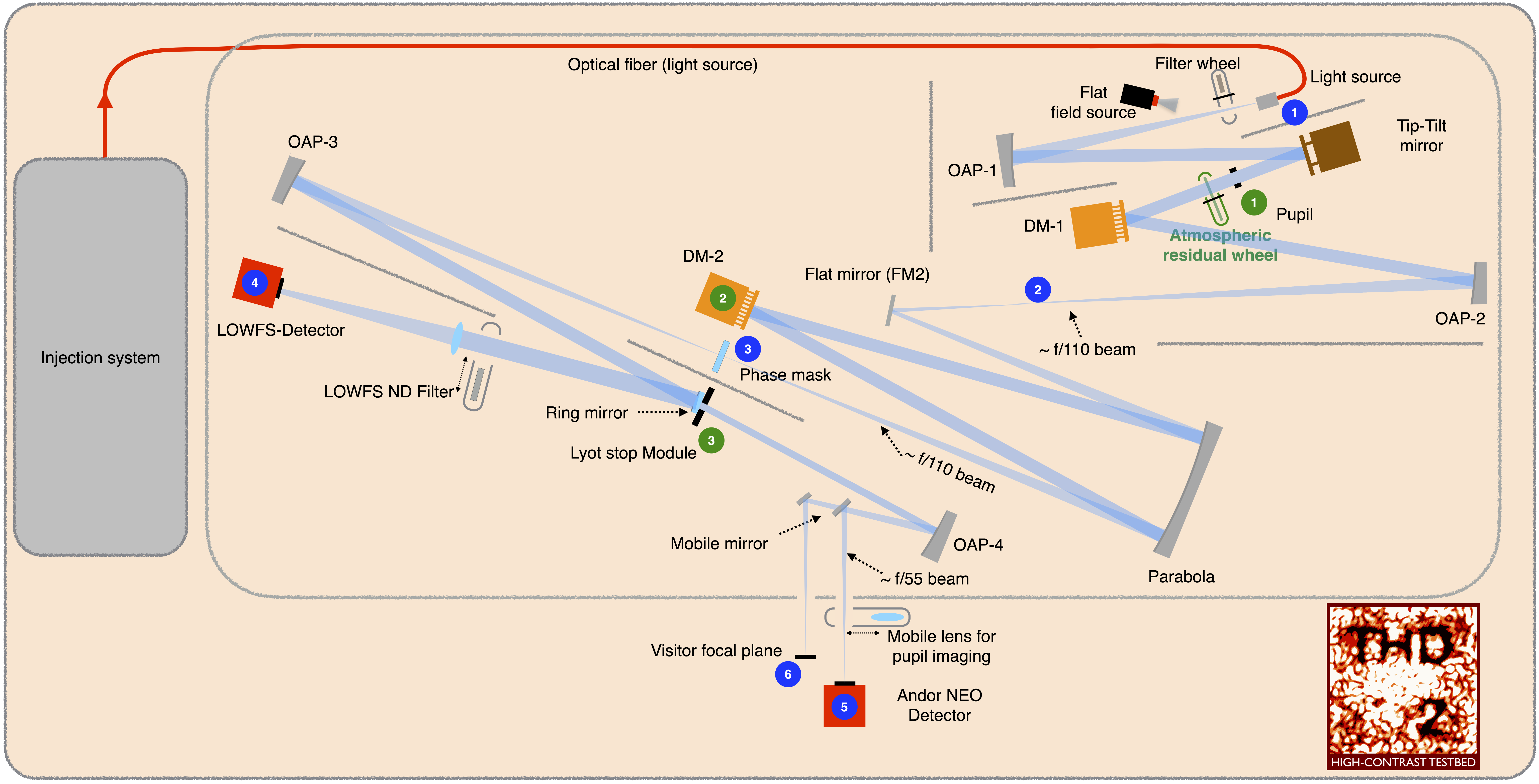}
    \caption{THD2 design showing the different components that could introduce polarization effects.}
    \label{fig:THD2_design}
\end{figure}

\begin{figure}[h]
\centering
    \includegraphics[width=\linewidth]{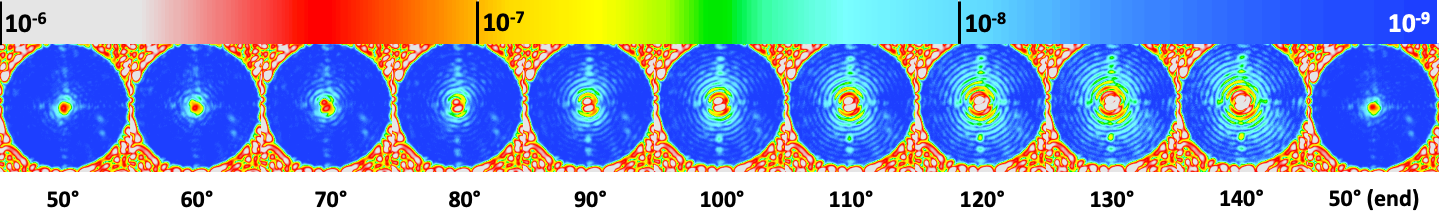 }
    \caption{Image contrast recorded on THD2 as a function of analyzer angles (placed in front of the camera). The dark dole (DH, best contrast) was created with the analyzer at a 50$^{\circ}$ orientation (mount angle with an arbitrary reference orientation with respect to the bench). The input polarization is arbitrary.}
    \label{fig:DH50to140deg}
\end{figure}

The THD2 bench is a unique testbed in Europe for high-contrast imaging. It has been developed at LESIA-Observatoire de Paris to test high-contrast imaging solutions in the visible and near-infrared at a contrast level of about $10^{-9}-10^{-8}$ in the context of exoplanet imaging for current as well as future telescopes, whether they are ground or space-based. Precise aberration correction for ground-based instruments has been tested on THD2\cite{Singh2019} and applied on SPHERE/VLT\cite{Potier22}. Experimental studies are also conducted in the context of optimal correction of space based instruments like the Roman Space Telescope\cite{Bailey2023} and the future Habitable World Observatory\cite{Gaudi2020_HWO}. While contrast levels of a few  $10^{-9}$ have been reached in this contex\cite{Potier20_THD2}, it was obtained filtering a single linear polarization at the entrance of the imaging camera. 
Departing from this initial final polarization state and without changing  the correction applied to the deformable mirrors (DM), we measured the performance for the different linear polarization states. We observed a degradation of the performance as shown in Fig. \ref{fig:DH50to140deg}. This demonstrates the existence and the impact of polarization differential aberrations. These images have been recorded with a laser centered at 785 nm and a Four Quadrant Phase Mask (FQPM)\cite{Rouan2000}.

The procedure applied to create the images shown in Fig. \ref{fig:DH50to140deg} is:
\begin{itemize}
\item Calculate a full dark hole (DH) correction using both DMs on THD2 bench with a linear analyzer placed in front of the camera oriented with an angle of 50$^{\circ}$ (mount angle, not related to any orientation on the bench). The DH correction is calculated based on pair-wise sensing and electric field correction as described in Potier et al. 2020\cite{Potier20_THD2} for a full DH using both DMs. No polarizer is place upstream in the bench and the polarization state of the entrance fiber is arbitrary.
\item Keep the obtained DM shapes at 50$^{\circ}$ for all the following measurements.
\item Set the reference tip-tilt position on the low-order aberration wave front sensing (LOWFS) loop for this analyzer orientation. 
\item Close the LOWFS loop during the rest of the measurements (note that the light used by the LOWFS is taken before the analyzer and is thus not affected by its orientation).
\item Record images of the DH from 50$^{\circ}$ to 140$^{\circ}$ and record a final image at 50$^{\circ}$ to make sure no drift occurred during the measurements.
\item Record off-axis images of the point-spread function (PSF) for each orientation to normalize the DH images.
\end{itemize}

The results in Fig. \ref{fig:DH50to140deg} show a clear degradation of the DH as we move away from the initial polarization state. The morphology of the DH images mostly mimic a tip-tilt error.

\begin{figure}[h]
\centering
    \includegraphics[width=0.35\linewidth]{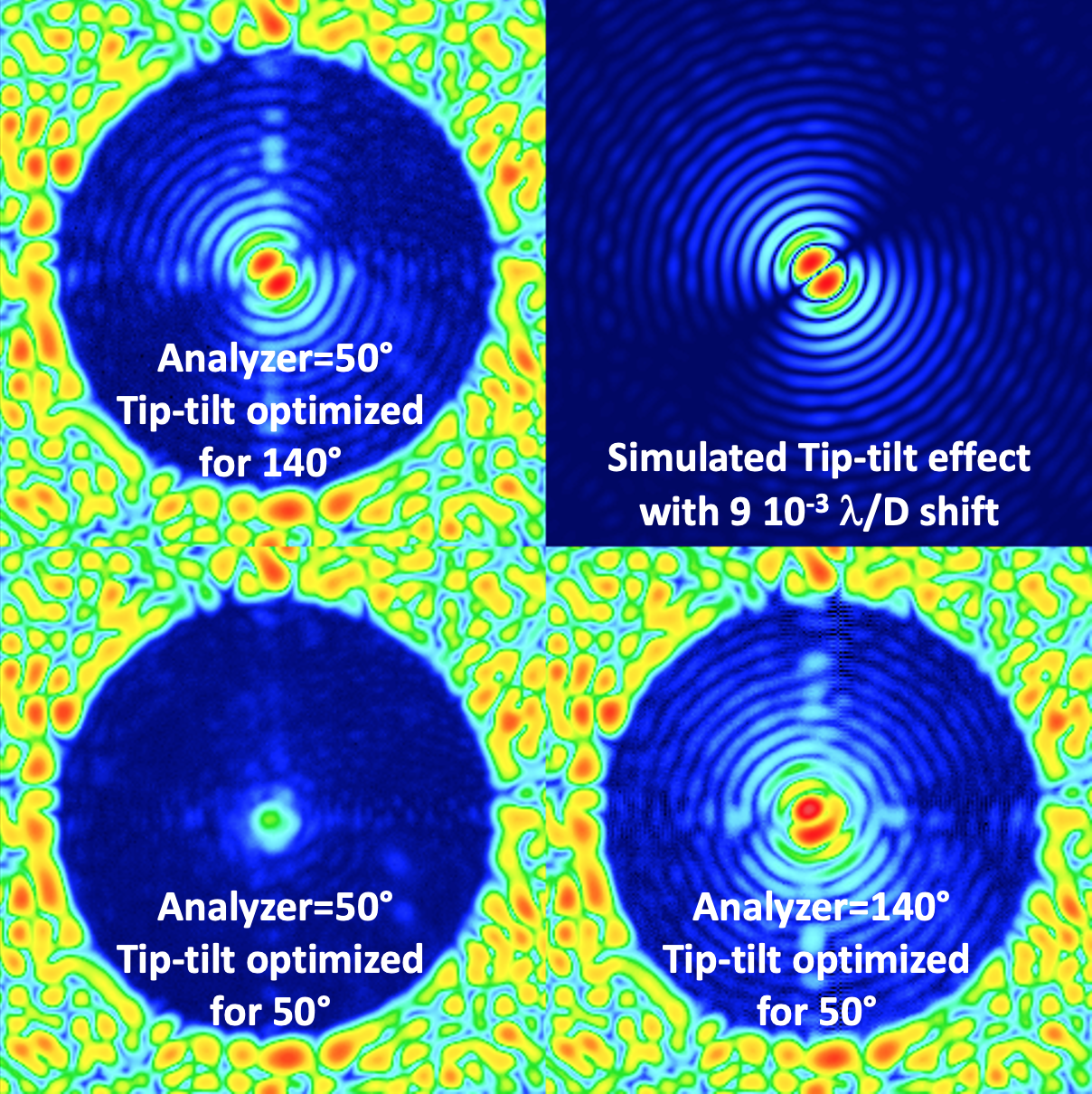}
    ~~
    \includegraphics[width=0.35\linewidth]{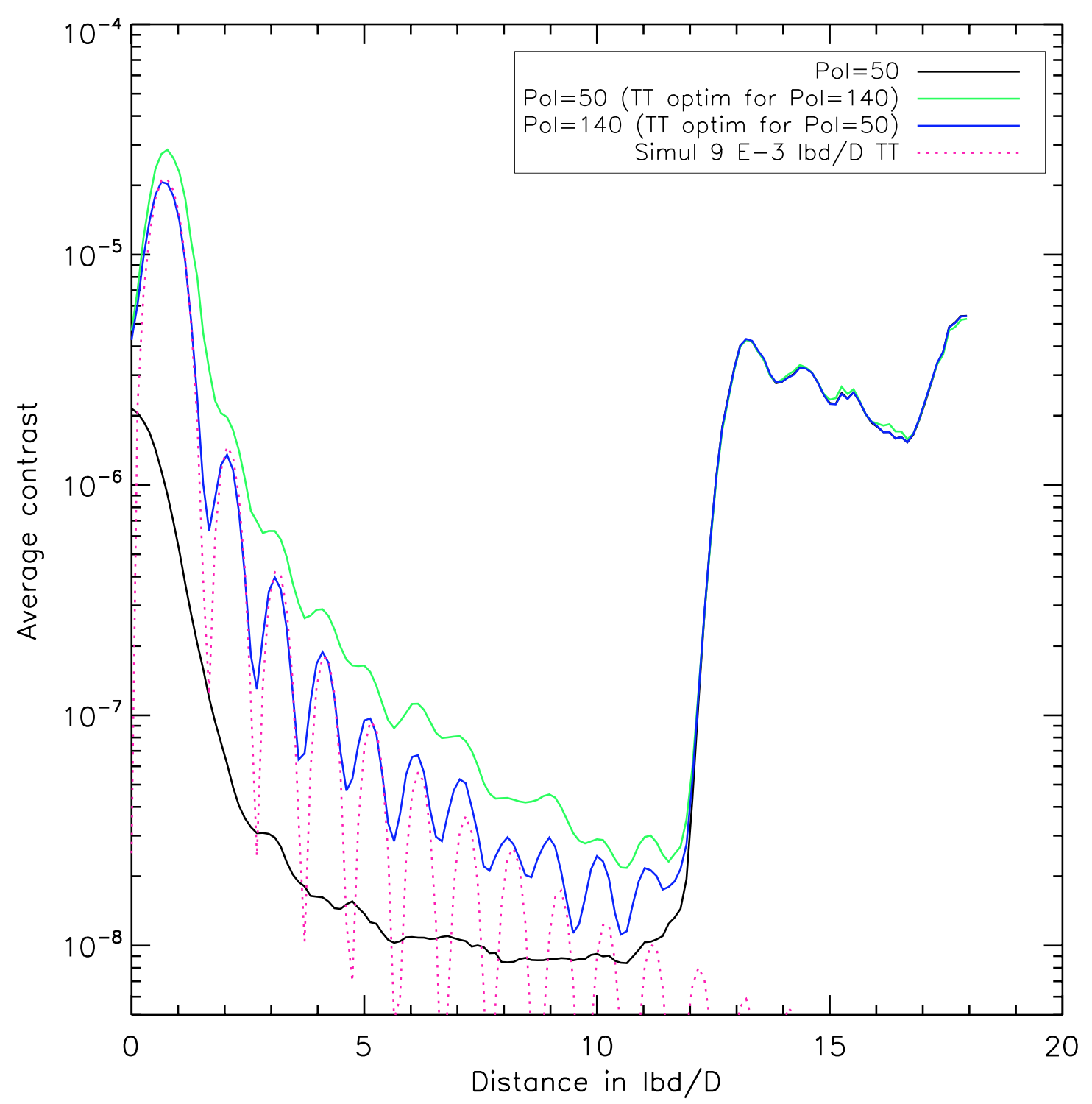}
    \caption{Bottom left: DH images for 50$^{\circ}$ and 140$^{\circ}$ analyzer angles (like in Fig.\ref{fig:DH50to140deg}). Top left: DH image for 50$^{\circ}$ analyzer angle using the 140$^{\circ}$-optimized LOWFS calibration and simulated image with 800 nm tip-tilt shift ($9 \cdot 10^{-3} \lambda/D$). Right: Radial profiles of the normalized intensity of the 4 images.}
    \label{fig:CompareDH50_DH140}
\end{figure}

We verified this effect by comparing the DH images at different analyzer angles with controlled tip-tilt aberrations in Fig. \ref{fig:CompareDH50_DH140}:
\begin{itemize}
    \item The DH recorded with analyzer=50$^{\circ}$ using 50$^{\circ}$-optimized DM shapes and 50$^{\circ}$-optimized LOWFS. This image corresponds to the first image on the left of Fig. \ref{fig:DH50to140deg}.
    \item The DH recorded with analyzer=140$^{\circ}$ using 50$^{\circ}$-optimized DM shapes and 50$^{\circ}$-optimized LOWFS. This image corresponds to the second to last image of Fig. \ref{fig:DH50to140deg}.
    \item The DH recorded with analyzer=140$^{\circ}$ using 50$^{\circ}$-optimized DM shapes and 140$^{\circ}$-optimized LOWFS.
    \item A simulated image with no aberration except a tip-tilt of $9 \cdot 10^{-3} \lambda/D=800\,$nm that could explain the shape of the previous image. 
\end{itemize}

From this first analysis, it seems that the PSF position at the coronagraph plane is different for each polarization state. It can have a strong impact on the performance of a coronagraph especially if it is very sensitive to tip-tilt as the FQPM used for these tests. 
Thus, we need to understand where this effect comes from to minimize its impact on our bench but also on future instruments.

\section{The Goos-H\"anchen and Imbert-Fedorov effects \label{sec:theory}}

\subsection{Overview\label{sec:overview_GH_IF}}

\begin{figure}[h]
\centering
    \includegraphics[width=0.5\linewidth]{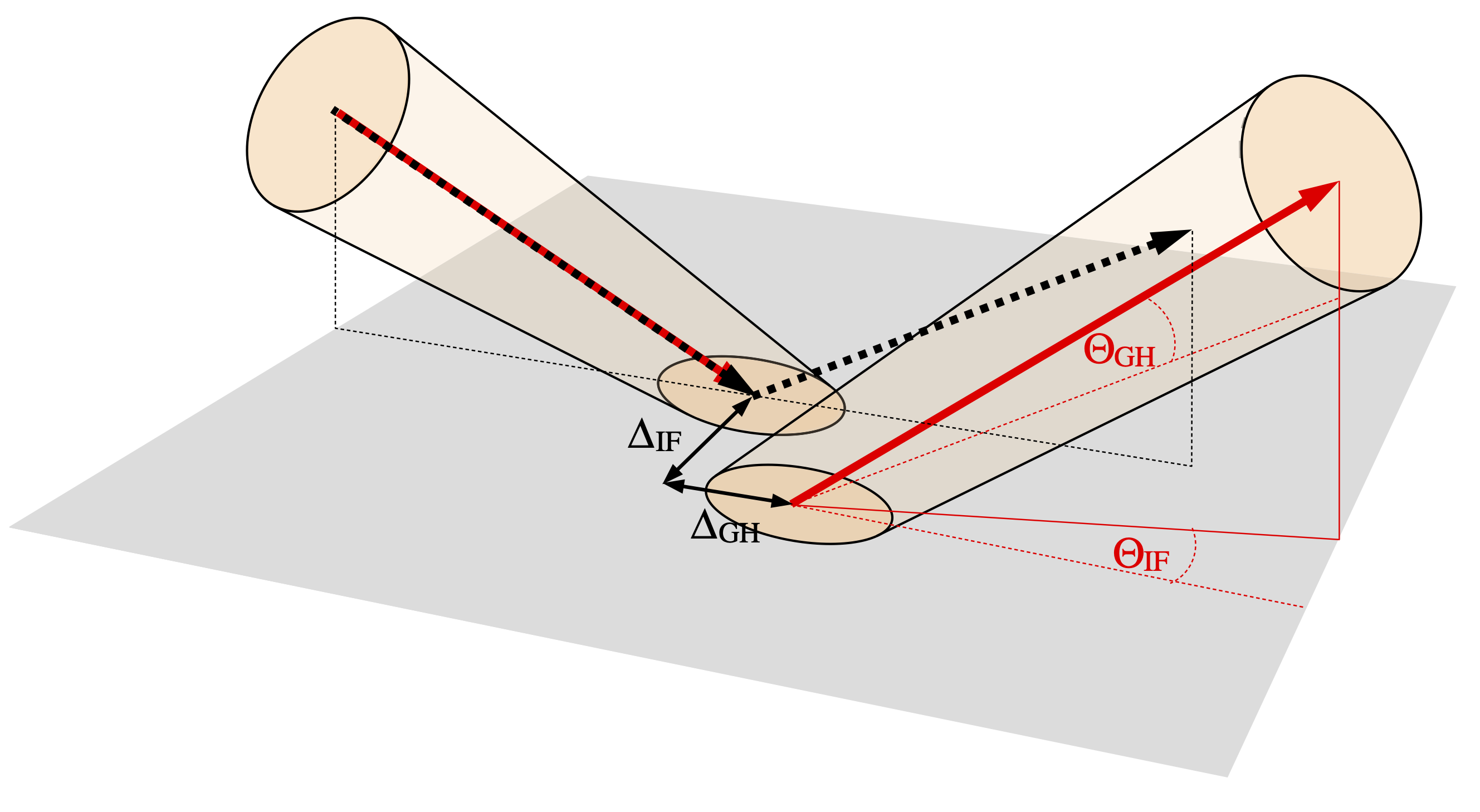}
    \caption{Scheme of the Goos-H\"anchen and Imbert-Fedorov effects. 
    The normal of the reflection plane surface is in the direction of the vector $\hat{n}$.  
    The incident plane is defined by the vectors $\hat{z}_i$ and $\hat{z}_r$ which represent the vectors of the incident and reflected beam in geometric optics, respectively (drawn in black here). As for $\hat{n}_r$, it is the observed reflected beam, respecting the GH and IF effects. The spatial shifts are noted by $\Delta_\text{GH}$ and $\Delta_\text{IF}$. The angular GH and IF shifts are also shown in red by the angular deviation between $\hat{z}_r$ and $\hat{n}_r$ \textit{in} the incident plane (GH) and also the angular deviation \textit{orthogonal} to the incident plane (IF). Figure from T\"oppel 2013\cite{Toppel_2013_GH&IF_shift}.}
    \label{fig:theory_GH&IF_shift}
\end{figure}

In geometric optics, light reflection and refraction at a plane dielectric interface is a basic, well-known process. The wave vectors of the incident and secondary waves are described by Snell's law and their polarization amplitudes are related by the Fresnel formulas \cite{born_wolf}. However, at wavelength scale it turns out four deviations are observed: a spatial and angular displacement \textit{in} the plane of incidence and \textit{normal} to the plane of incidence, represented in Fig. \ref{fig:theory_GH&IF_shift}.
The Goos-H\"anchen (GH) effect corresponds to the \textit{spatial and angular} displacement \textit{in} the plane of incidence of the beam. The eigenvectors of these displacements correspond to the linear polarizations $p$ or $s$-polarization, i. e. TM or TE-polarization.
Regarding the Imbert-Fedorov (IF) effect, the \textit{spatial and angular} displacements are \textit{normal} to the plane of incidence.
The eigenvectors for the spatial displacements correspond to the circular polarizations $left$ and $right$. As for the angular deviation, the eigenvectors correspond to the combinations of linear polarizations oriented at $45 \degree$  to the linear polarizations $p$ and $s$\cite{Bliokh2013,Goswami}.
Both the GH and IF effects depend on the refractive index, the incident angle and the wavelength. The IF effect also depends on the nature of the beam (Gaussian, Bessel...).

The GH and IF effects could be responsible of the optical aberrations observed in Sect. \ref{sec:intro} as they are polarization dependent. These effects have been studied in optics since Artmann 1948 but dozens of publications have been published on this matter between $2000$ and $2020$, covering theoretical and experimental perspectives, with different formalisms, based on wave optics and quantum mechanics. The IF effect is also known under other names, as the spin Hall effect of light. To date, questions on these shifts still remain. In spite of this interest in optical research, the effects have only recently been observed\cite{Schmid2018} and studied\cite{VanHolstein23} in astrophysics, probably due to the improved sensitivity of instruments.

\subsection{Analytical expressions \label{sec:expressions_GH&IF_effects}}

First, we calculate the Goos-H\"anchen and Imbert-Fedorov effects on the THD2 bench for metallic mirrors which have a complex refractive index. Here, we will use the analytical expressions given in Aiello et al. 2009 and Aiello 2012\cite{Aiello2009,Aiello2012}. The first paper has the main advantage to give a nice explicit formalism to express the GH and IF effects in terms of the incident angle $\theta$ and the relative permittivity $\varepsilon=\varepsilon_r+i\,\varepsilon_i$ which is related to the complex refractive index $n$ by the expression $n^2 = \varepsilon_r + i\, \varepsilon_i$. These expressions depend on the nature of the beam, which they assume to be gaussian.

\begin{figure}[!h] \centering
\includegraphics[angle=0,width=0.5\linewidth,
]{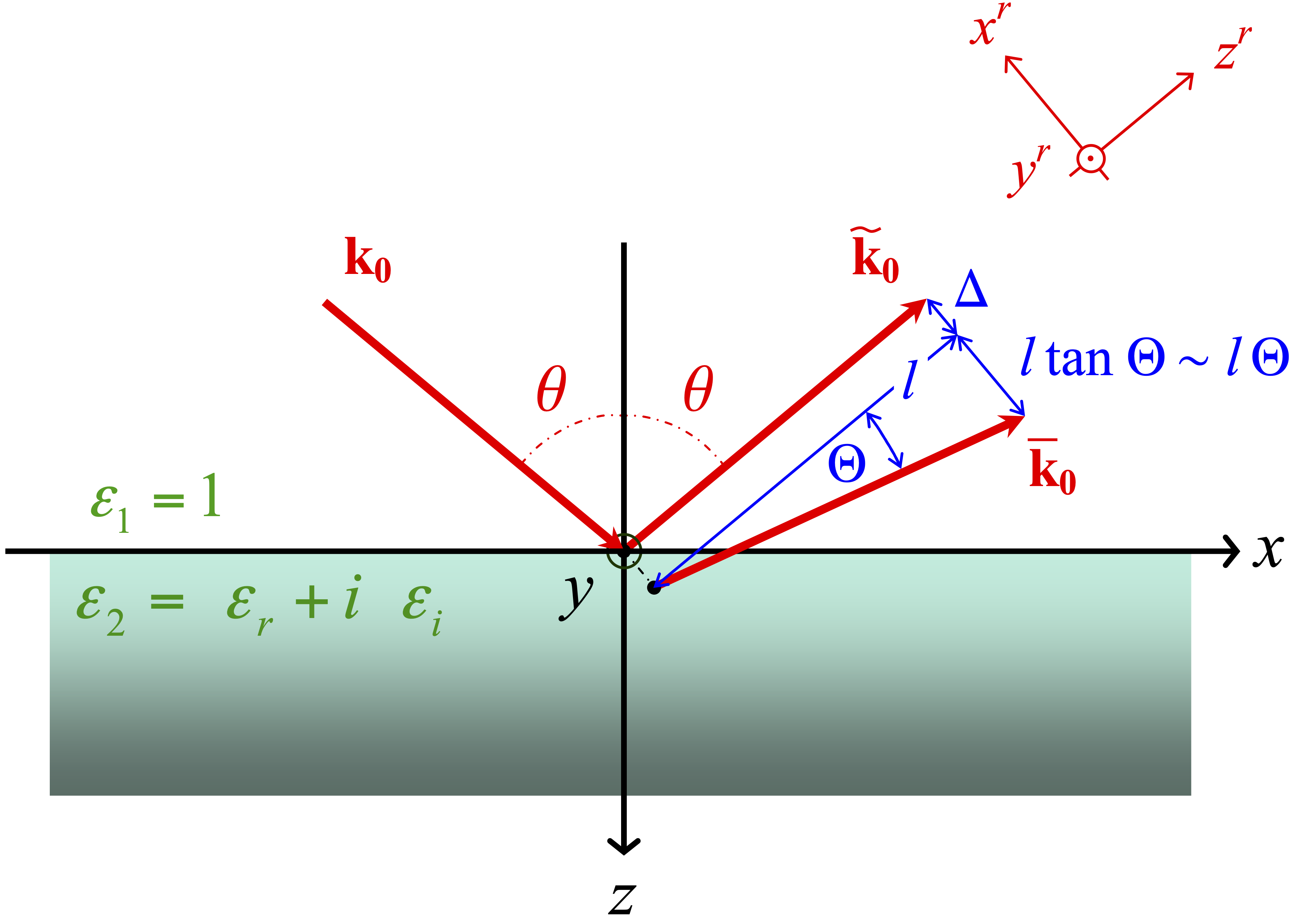}
\caption{\label{fig:Aiello_scheme_interface_GH_IF}
Scheme of the beam reflection at the plane interface (x-axis) in the incident plane ($O_{xy}$). The central wave vectors of the incident beam is  $\mathbf{k}_0$, while the vectors $\widetilde{\mathbf{k}}_0$ and $\overline{\mathbf{k}}_0$ are the central wave vectors of the reflected beam as predicted by geometrical and wave optics, respectively. The Goos-H\"anchen spatial and angular displacements are noted by $\Delta$ and $\Theta$. For a camera located at a given distance $l$ from the plane interface, the angular deviation is seen as a spatial deviation $l \Theta$.  Figure from Aiello et al. 2009\cite{Aiello2009}.}
\end{figure}

The Goos-H\"anchen and Imbert-Fedorov effects are both characterized by a spatial and angular shift, noted respectively by $\Delta$ and $\Theta$, due to the presence of a plane interface. In the case of the GH effect, the angular and spatial shifts, as well as the notations, are illustrated in Fig. \ref{fig:Aiello_scheme_interface_GH_IF}. Medium $1$ is air, while medium $2$ a dielectric or a metal. Consequently, the dielectric constants are $\varepsilon_1 = 1$ and $\varepsilon_2 = \varepsilon_r + i\,\varepsilon_i$, with $\varepsilon_r > 1 $ if medium $2$ is a dielectric or $\varepsilon_r < 0 $ if it is a metal.

The global beam shift can be expressed at a distance $l$ of the interface as:
\begin{equation}\label{eq:theory_global_shift}
\delta_\chi(l)  \;=\; \Delta_\chi \,+\, l \, \Theta_\chi \,, 
\end{equation}
by considering Fig. \ref{fig:Aiello_scheme_interface_GH_IF} and assuming $\Theta_\chi \ll 1$.

The expression of the spatial $\Delta_\chi$ and the angular $\Theta_\chi$ displacements, where $\chi$ represents the $p$ or $s$-polarization are reported by Aiello et al. 2009\cite{Aiello2009} as:
\begin{equation}\label{eq:spa_ang}
\Delta_\chi \;=\;  \frac{\lambda_0}{2 \pi} \, \text{Im} \left[ D_\chi \right] \qquad \textrm{and} \qquad \Theta_\chi \;=\;  - \, \frac{1}{2} \left({\frac{\lambda_0}{\pi w_0}}\right)^2 \, \text{Re} \left[ D_\chi \right],
\end{equation} 
where $\lambda_0$ is the central wavelength, $\theta$ the angle of incidence, $w_0$ the waist of the incident beam, and the coefficient $D_\chi$ detailed now.
\subsubsection{Goos-H\"anchen shift}

In the GH case, the coefficient $D_\chi$ is equal to
\begin{equation}\label{eq:D_shift_GH}
D_\chi \;=\;   \frac{\partial \ln r_\chi}{\partial \theta}\,, 
\end{equation}
which gives, by using the expressions of the Fresnel reflection coefficients\cite{born_wolf}:
\begin{align}
D_p  \;&=\;   -\, \frac{2 \sin \theta}{\sqrt{\left(\varepsilon_r - \sin^2 \theta \right) + i\, \varepsilon_i}} \, \frac{\varepsilon_i^2 + \varepsilon_r\left( 1 - \varepsilon_r\right)  + i\, \varepsilon_i \left( 1 - 2 \varepsilon_r\right)}{
  \left(\varepsilon_r - \sin^2 \theta \right)  + \left(\varepsilon_i^2 -\varepsilon_r^2 \right)\cos^2 \theta  + i\, \varepsilon_i \left( 1 - 2 \varepsilon_r \cos^2 \theta \right) }  \\ 
D_s  \;&=\;  \frac{2 \sin \theta}{\sqrt{\left(\varepsilon_r - \sin^2 \theta \right) + i\, \varepsilon_i}} \,. \label{eq:GH_DP_DS}
\end{align}
These final expressions were given in Aiello et al. 2009\cite{Aiello2009}, but with a typo in their paper. We confirmed with them that they forgot a ”-” in the expression of $D_p$ in their Eq. (8).

\subsubsection{Imbert-Fedorov shift}

As for the IF case, Aiello 2012\cite{Aiello2012} reports coefficients $D_p$ and $D_s$ equal to
\begin{equation}\label{eq:D_shift_IF}
D_p \;=\;   i \, \frac{r_p + r_s}{r_\chi} \, \cot \theta \quad \text{and} \quad D_s \;=\; -\, i \, \frac{r_p + r_s}{r_\chi} \, \cot \theta \,,
\end{equation}
from which can be found as final expression of the IF shift
\begin{equation}\label{eq:IF_Dp_eps}
D_p \;=\; 2\,i\,\cot\theta
\,\times\,
\frac{\sin^2\theta} 
{\left({\sin^2\theta \,-\, \cos\theta\, \sqrt{\varepsilon_r\,-\,\sin^2\theta\,+\,i\,\varepsilon_i}}\right)}\,,
\end{equation} 
\medskip
and
\begin{equation}\label{eq:theory_Ds_IF_eps} 
D_s     \;=\; -\,2\,i\,\cot\theta
\,\times\,
\frac{\sin^2\theta\,} 
{\left({\sin^2\theta \,+\, \cos\theta\, \sqrt{\varepsilon_r\,-\,\sin^2\theta\,+\,i\,\varepsilon_i}}\right)}\,.
\end{equation}

\subsection{Application to the THD2 bench: Theoretical GH and IF shift values} \label{sec:theory_application_thd2}

In this section, we calculate the theoretical Goos-H\"anchen and Imbert-Fedorov effects for each mirror of the THD2 bench assuming all mirrors (OAPs, DMs, tipt-tilt mirror and final fold mirror in front of the camera) are recovered by a silver layer (neglecting the protected layer of silver and the fact that the DMs are aluminum coated). The incident angles on the bench are ranging from $3.16\degree$ to $14.99\degree$. The relative permittivity for silver is taken as $\varepsilon = -19.69 + 1.24\,i$ at $639$ nm, $\varepsilon = -22.45 + 1.33\,i$ at $705$ nm and $\varepsilon = -25.89 + 1.46\,i$ at $785$ nm\cite{Palik1997}.

First, we calculate the effects on the flat mirror FM2, as it has the greatest angle of incidence, $i = 14.99\degree$. As FM2 is located in the converging beam of the OAP2, at a $254$-mm distance from the focal plane, the distance $l$ to turn the angular shift $\Theta$ into the spatial shift $l\Theta$ is $254$ mm. Since OAP2 has a $900$-mm focal distance, which is the same as the focal distance of the parabola preceeding the coronagraph, there is no optical magnification to compute the polarization effects caused by FM2 at the coronagraph focal plane.
We present all the values calculated in Table \ref{tab:GH&IF_predicted_values_THD_2}. In this table, we gather the predicted Goos-H\"anchen and Imbert-Fedorov spatial ($\Delta$) and angular ($\Theta$) shifts in percentage of the total shift values calculated either for the final focal plane (full bench) or for the coronagraph focal plane. Shift values at this latter plane is also given in nm to compare with the indirect measurement done in Sect. \ref{sec:intro}. The parameters $i$, $\gamma$ and $l$ represent the incident angle of the beam on the medium, the optical magnification at coronagraph (or camera) focal plane used to compute the predicted shifts and the distance used to turn the angular shift $\Theta$ into the spatial shift $l \Theta$. For collimated beams, the distance $l$ is the focal distance of the last OAP/parabola before the frame of interest (i. e. parabola for the coronagraph and OAP4 for the camera). Concerning the converging beams, as for FM2, it is the distance between the mirror (for example FM2) and the focal plane that must be taken into account. The optical magnification must also be used if the focal distance of the OAP/parabola preceding the frame of interest is different of the one in which the mirror is located. Concerning the spatial deviations, only the mirrors that converge the beam and FM2, which is in a diverging beam, have to be taken into account. 

\begin{figure}
    \centering
    \includegraphics[width=\linewidth]{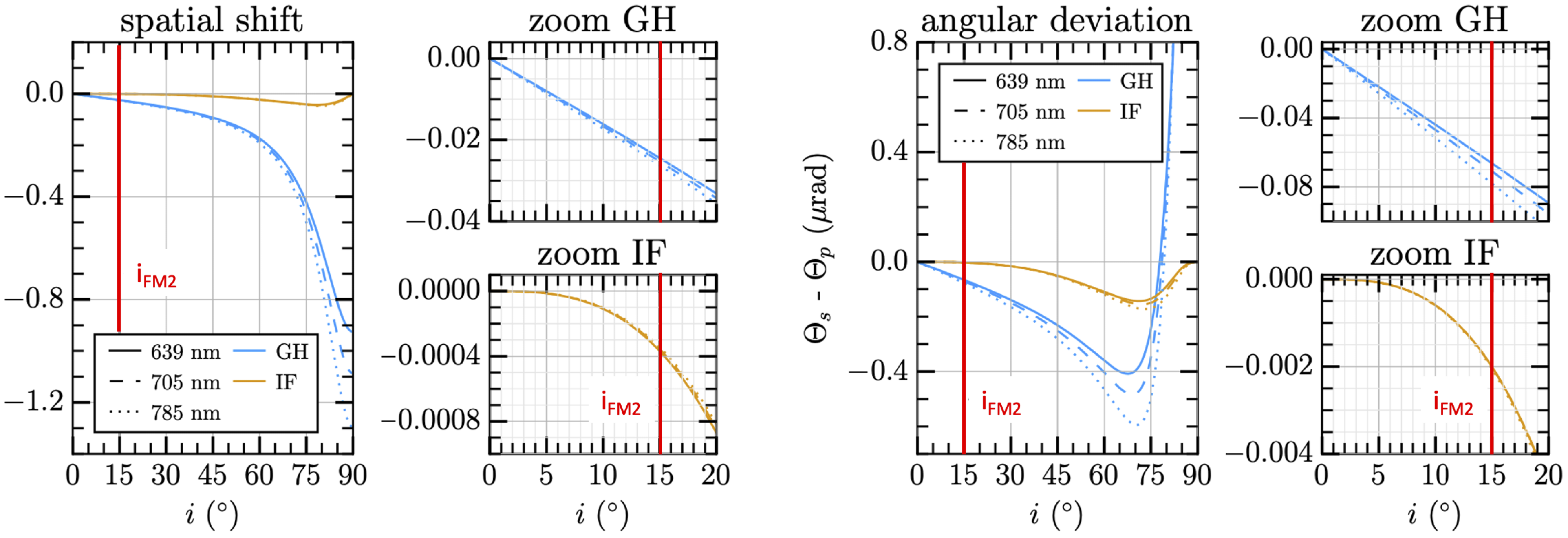}
    \caption{
    The GH and IF spatial shift (\textit{left}) and angular (\textit{right}) deviation $s-p$ on the THD2 bench as a function of the incident angle $i$ on a silver coating mirror. The red vertical line at $\sim 15\degree$ represent the incident angle on the FM2, which has the greatest incident angle on the THD2 bench (FM2 is the flat mirror located after the OAP2, see Fig. \ref{fig:THD2_design}).}
    \label{fig:GHshift_Ag_theory}
\end{figure}

According to Fig. \ref{fig:GHshift_Ag_theory}, both the spatial and angular Goos-H\"anchen effects are significantly greater than spatial and angular Imbert-Fedorov effect, respectively of a factor $70$ and $35$ for FM2 ($i\sim15\degree$). Unlike the Goos-H\"anchen effects which look linear for $i$ smaller than $15\degree$, the Imbert-Fedorov effects are non linear. Furthermore, the dependence on the wavelength seems negligible for most of the shifts, except maybe for the Goos-H\"anchen angular deviation.

\begin{table}[h!]
\resizebox{\textwidth}{!}{
    \centering 
    \begin{tabular}{l|c|c|c|c|c|c|c|c||c|c|c|c|c} 
        \hline \hline
         Medium & OAP1 & TT & DM1 & OAP2 & FM2 & Parab. & DM2 & Parab. & OAP3 & OAP4 & Fold & Total & Unit \\
        \hline\hline
        $i$ & $3.26$& $6.52$& $11.52$& $3.23$& $14.99$ & $3.31$& $6.55$& $3.31$& $3.29$& $3.16$ & $12.5$ & $-$ & ($\degree$)\\
        \hline
        $l$ coro.& $900$& $900$& $900$& $900$& $254$ & $900$& $900$& $900$ & $-$& $-$& $-$ & $-$ & (mm) \\
        $l$ cam.  & $500$& $500$& $500$& $500$& $254$ & $500$& $500$& $500$ & $500$& $500$& $347.5$ & $-$ & (mm)\\
        \hline
        $\gamma$ coro. & $1$& $1$& $1$& $1$& $1$ & $1$& $1$& $1$& $-$& $-$ & $-$ & $-$ & (\O)\\
        $\gamma$ cam. & $1$& $1$& $1$& $1$& $0.56$ & $1$& $1$& $1$& $1$& $1$ & $1$ & $-$ & (\O)\\
        \hline\hline
        $\Delta_\text{GH}$ & \multicolumn{12}{|c}{\GHcolor{Spatial Goos-H\"anchen shifts}} \\ 
        \hline\hline 
        full bench &$-$&$-$&$-$&$10$& $28$ &$-$&$-$&$11$&$-$&$10$& $41$ & $100$ & (\%)\\ 
        coro. &$-$&$-$&$-$&$15$& $70$ &$-$&$-$&$15$&$-$&$-$& $-$ & $100$ & (\%)\\ 
        &$-$&$-$&$-$&$5.53$& $2.98$ &$-$&$-$&$5.62$&$-$&$-$& $-$ & $14.13$ & (nm)  \\ 
        \hline\hline
        $\Theta_\text{GH}$ & \multicolumn{12}{|c}{\GHcolor{Angular Goos-H\"anchen deviations}} \\
        \hline\hline
        full bench & $6$ & $11$ & $20$ & $6$ & $7$ & $6$ & $11$ & $6$ & $6$ & $6$ & $15$ & $100$ & (\%) \\ 
        coro. & $8$ & $16$ & $28$ & $8$ & $10$ & $8$ & $16$ & $8$ & $-$ & $-$ & $-$ & $100$ & (\%) \\ 
          & $4.77$ & $9.53$ & $16.94$ & $4.68$ & $6.30$ & $4.77$ & $9.53$ & $4.77$ & $-$ & $-$ & $-$ & $61.29$ & (nm)\\ \hline
         & \multicolumn{12}{|c}{\GHcolor{ Total Goos-H\"anchen effects (Horizontal shift on THD2)}} \\ 
         \hline
          & $4.77$ & $9.53$ & $16.94$ & $10.21$ & $9.28$ & $4.77$ & $9.53$ & $10.39$ & $-$ & $-$ & $-$ & $75.42$ & (nm)\\ 
        \hline\hline 
        $\Delta_\text{IF}$ & \multicolumn{12}{|c}{\IFcolor{Spatial Imbert-Fedorov shifts}} \\ 
        \hline\hline
        full bench &$-$&$-$&$-$&$1$& $47$ &$-$&$-$&$1$&$-$&$-$& $49$ & $100$ & (\%) \\ 
        coro. &$-$&$-$&$-$&$1$& $98$ &$-$&$-$&$1$&$-$&$-$& $-$ & $100$ & (\%) \\ 
         &$-$&$-$&$-$&$0.00$& $0.34$ &$-$&$-$&$0.00$&$-$&$$& $$ & $0.35$ & (nm)  \\ 
        \hline\hline
        $\Theta_\text{IF}$ & \multicolumn{12}{|c}{\IFcolor{Angular Imbert-Fedorov deviations}} \\
        \hline\hline
        full bench & $1$ & $6$ & $33$ & $1$ & $21$ & $1$ & $6$ & $1$ & $1$ & $1$ & $29$ & $100$ & (\%) \\ 
        coro. & $1$ & $9$ & $48$ & $1$ & $30$ & $1$ & $9$ & $1$ & $-$ & $-$ & $-$ & $100$ & (\%) \\ 
         & $0.01$ & $0.05$ & $0.27$ & $0.01$ & $0.16$ & $0.01$ & $0.05$ & $0.01$ & $$ & $$ & $$ & $0.55$ & (nm) \\ 
         \hline
          & \multicolumn{12}{|c}{\IFcolor{Total Imbert-Fedorov effects (Vertical shift on THD2)}} \\
        \hline
        & $0.01$ & $0.05$ & $0.50$ & $0.01$ & $0.16$ & $0.01$ & $0.05$ & $0.01$ & $$ & $$ & $$ & $0.90$ & (nm) \\ 
        \hline
         \hline
    \end{tabular}}
    \smallskip
    \caption{Predicted Goos-H\"anchen and Imbert-Fedorov spatial ($\Delta$) and angular ($\Theta$) shifts in percentage of the total values calculated at the final focal plane values (full bench) or at the coronagraph focal plane. Shift values at this latter plane are also given in nm.
    The parameters $i$, $\gamma$ and $l$ represent the incident angle of the beam on the medium, the optical magnification at coronagraph (or camera) focal plane used to compute the predicted shifts and the distance used to turn the angular shift $\Theta$ into the spatial shift $l \Theta$. 
    The double vertical line indicates the position of the coronagraph, between the second reflection on the parabola and OAP3.
        } \label{tab:GH&IF_predicted_values_THD_2}
\end{table}

Note that the predicted values are derived from analytical expressions assuming a \textit{gaussian} beam. On the THD2 bench, the beam is not gaussian, but a truncated gaussian beam starting after the pupil plane. The values may still be approximately relevant asuming a waist ($w_0$) for the calculation equal to the FWHM of the beam at the focal plane.

\section{GH and IF shift measurements on THD2}

\subsection{Methodology to measure amplified GH and IF effects}

\begin{figure}[h!]
\centering
    \includegraphics[width=0.7\linewidth]{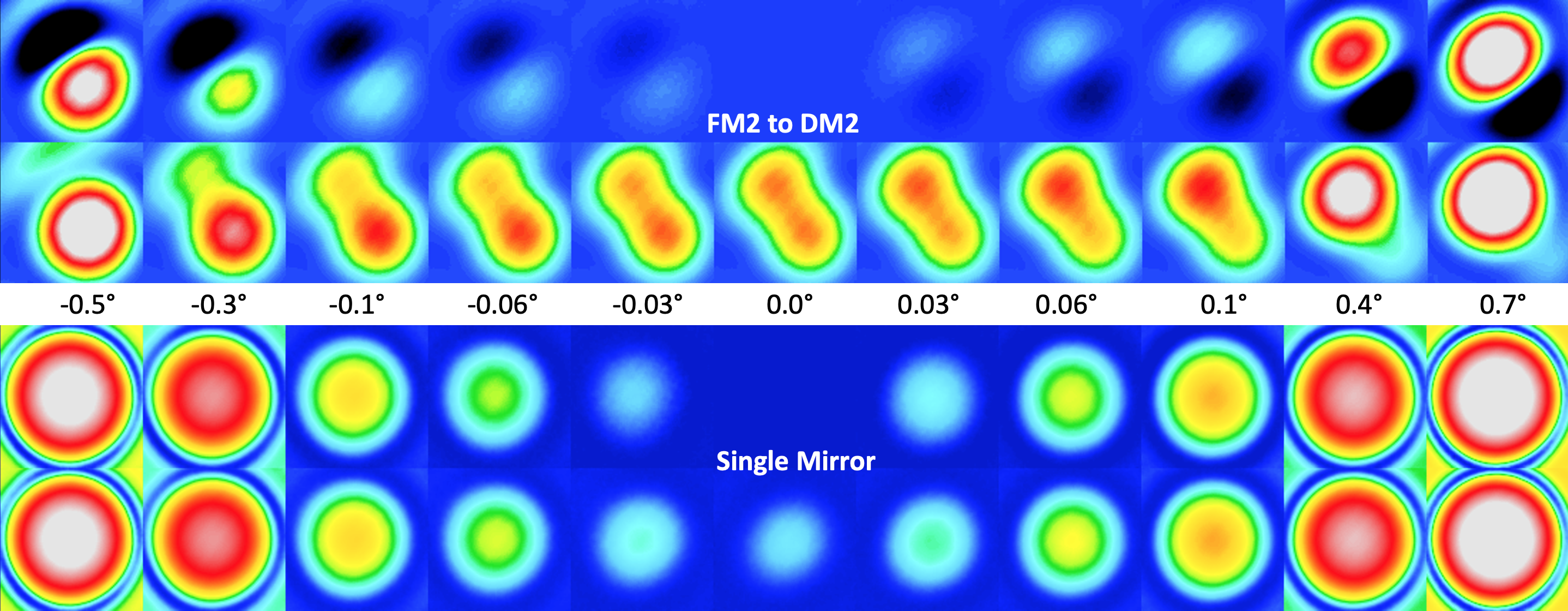}
    \caption{Images recorded around cross-polarization for the case of a simple mirror measurement (2 bottom rows) and for the optics including FM2, the first reflection on our parabola and DM2 (2 top rows). First and third rows show the image after subtracting the best attenuation image.}
    \label{fig:ImagesGHIF_measurements}
\end{figure}

The theoretical values in Table \ref{tab:GH&IF_predicted_values_THD_2} are more than ten times smaller than the values of the shift that was indirectly measured in Sect. \ref{sec:intro}. To better understand if this is a problem of normalization because the THD2 beam is not Gaussian shaped, we decided to better characterize the effects of polarization on THD2. The purpose is to detect the spatial and angular Goos-H\"anchen and Imbert-Fedorov shifts for different elements of the bench and compare them to the theoretical values.

The shifts estimated in Sect. \ref{sec:theory_application_thd2} or measured in Sect. \ref{sec:intro} are much smaller than the pixel size: 800 nm versus 6500 nm pixel size. To measure them, we decided to apply the weak measurement method described in Goswani et al. 2014\cite{Goswami}. The methodology is to measure the position of the beam reflected from a mirror (or several mirrors) located between a polarizer and an analyzer close to  the cross-polarization angle. We remove the coronagraph from the bench when we record these data. The weak measurement allows, when the polarization direction of the polarizer is nearly orthogonal to the analyzer direction, to see an amplification of the Goos-H\"anchen and Imbert-Fedorov shifts by a factor proportional to $\frac{1}{\omega-\omega_{\perp}}$ for small angles $\omega-\omega_{\perp}$\cite{Goswami}. We define experimentally $\omega_{\perp}$ by finding one of the two angles that allow to minimize the intensity transmission of the PSF. Our precision to find this cross polarization angle $\omega_{\perp}$ is about 0.01$\degree$  with typical leakage through the polarizer-analyzer below $10^{-6}$.

The amplification function $g$ we must fit is given by:
\begin{equation}\label{eq:fit_g}
g : \omega-\omega_{\perp} \;\mapsto\; \frac{q}{\omega-\omega_{\perp}} \,+\, d_2 \,,
\end{equation}
where $q$ is a positive or negative coefficient, and $d_2$ a constant to correct for the mean position value.

We measure the effects when the first polarizer selects either $p$ or $s$ polarization state. We record the images for a series of analyzer angle position spanned from 0$\degree$ to 360$\degree$ sampling every 10$\degree$ for the angles that are far from cross polarization and improving the sampling as we are getting close to $\omega_{\perp}$ by decreasing gradually the step angles to 0.01$\degree$ around cross-polarization. The detector used for these measurements has a small pixel size (2.2 $\mu$m) and we introduced an optical magnification at the output of the bench to increased the PSF size to about $\lambda/D=$355 pixels at 640 nm.

\begin{figure}[h!]
\centering
    \includegraphics[width=0.5\linewidth]{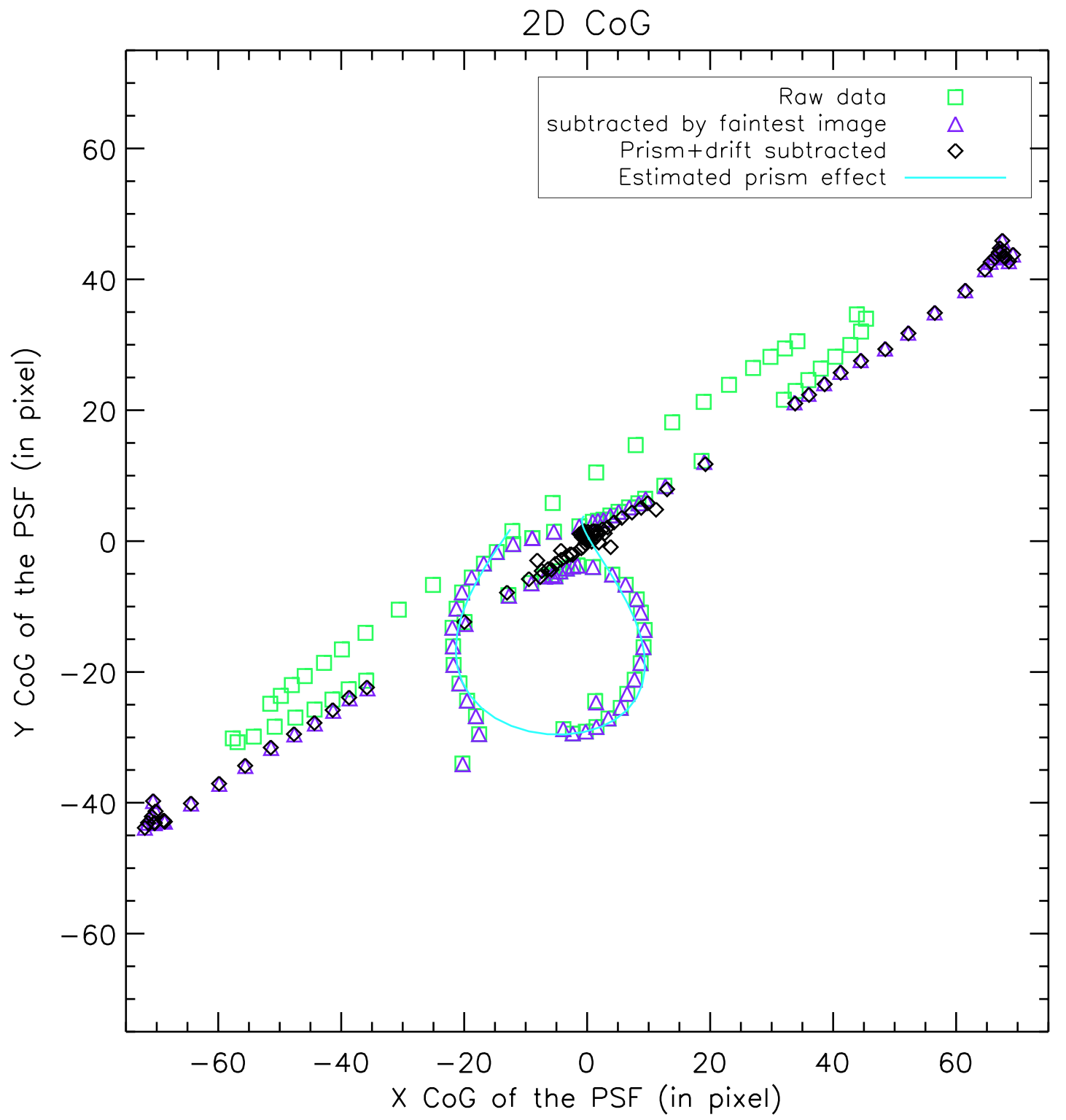}
    \caption{Example of center of gravities (CoG) for all measured angles as a function of pixel position on the detector. The example corresponds to the CoG of the top row images of Fig. \ref{fig:ImagesGHIF_measurements}. The CoG are given for the images without (green squares) and with subtracted best attenuation image (violet triangles), and prism subtracted CoG (black diamonds). The blue line shows the estimation of both the fitted prismatic effect of the analyzer and the temporal drift.}
    \label{fig:2D_cog}
\end{figure}

To measure the small motion of the beam ($\approx 10^{-2} \lambda/D$), we use the method of a weighted center of gravity (CoG)\cite{Nicolle2004} to minimize the noise effect on the estimation. The example of a CoG is shown in Fig. \ref{fig:2D_cog} where we can see different movements:
\begin{enumerate}
    \item One movement due to the complete rotation of the polarizer that is not a perfectly parallel substrate (period of $360\degree$). The small prismatic shape of the polarizer rotates the image on 360°.
    \item One effect that depends on the drifting of the beam as a function of the mechanical/optical variation probably related to temperature changes.
    \item The expected movement of $\frac{1}{\omega-\omega_{\perp}}$ on each axis.
\end{enumerate}

\subsection{Data processing of amplified measurements}

To suppress the drift due to mechanical/optical variations, the first and last image of each series is recorded under the same configuration, thus rotating the analyzer by 360$\degree$. Then, we assume the drift to be proportional with time, which is an empirical approximation that seems to be enough to give us an estimation of the movement.

We reduce the study of the 2D-displacements of the beam into two 1D-displacements (in X and Y), to quantify the additional shifts (1) and (2). 
Thereby, we fit the $x$ and $y$-positions of the beam by the function $f$ defined as follows:
\begin{equation}\label{eq:fit_f}
f : (\omega,t) \;\mapsto\; a\,\sin(\omega+\phi_1) +\,b\,\omega(t)\,+\,d_1 \,,
\end{equation}
where
\begin{itemize}
    \item $\omega$ is the angle of rotation of the analyzer. Here $\omega$ spans from 0$\degree$ to 360 $\degree$. 
    \item $a\,\sin(\omega+\phi_1$) corresponds to the shift due to the rotation of the prism ($360\degree$-periodic);
    \item $b\,\omega(t)$ takes into account the temporal drift present on THD2, as we took about the same amount of time to measure each image;
    \item $d_1$ is the mean position of the beam.
    \item $a,\,d_1\,b,\,d_1$ are parameters of the fit.  
\end{itemize}

This first fit to suppress the additional shifts (1) and (2) is done on all $\omega$ angles except the ones that are too close to the amplification effects, i.e. for angles $\|\omega-\omega_{\perp}\|> 2\degree$.

\begin{figure}
\centering
    \includegraphics[width=0.9\linewidth]{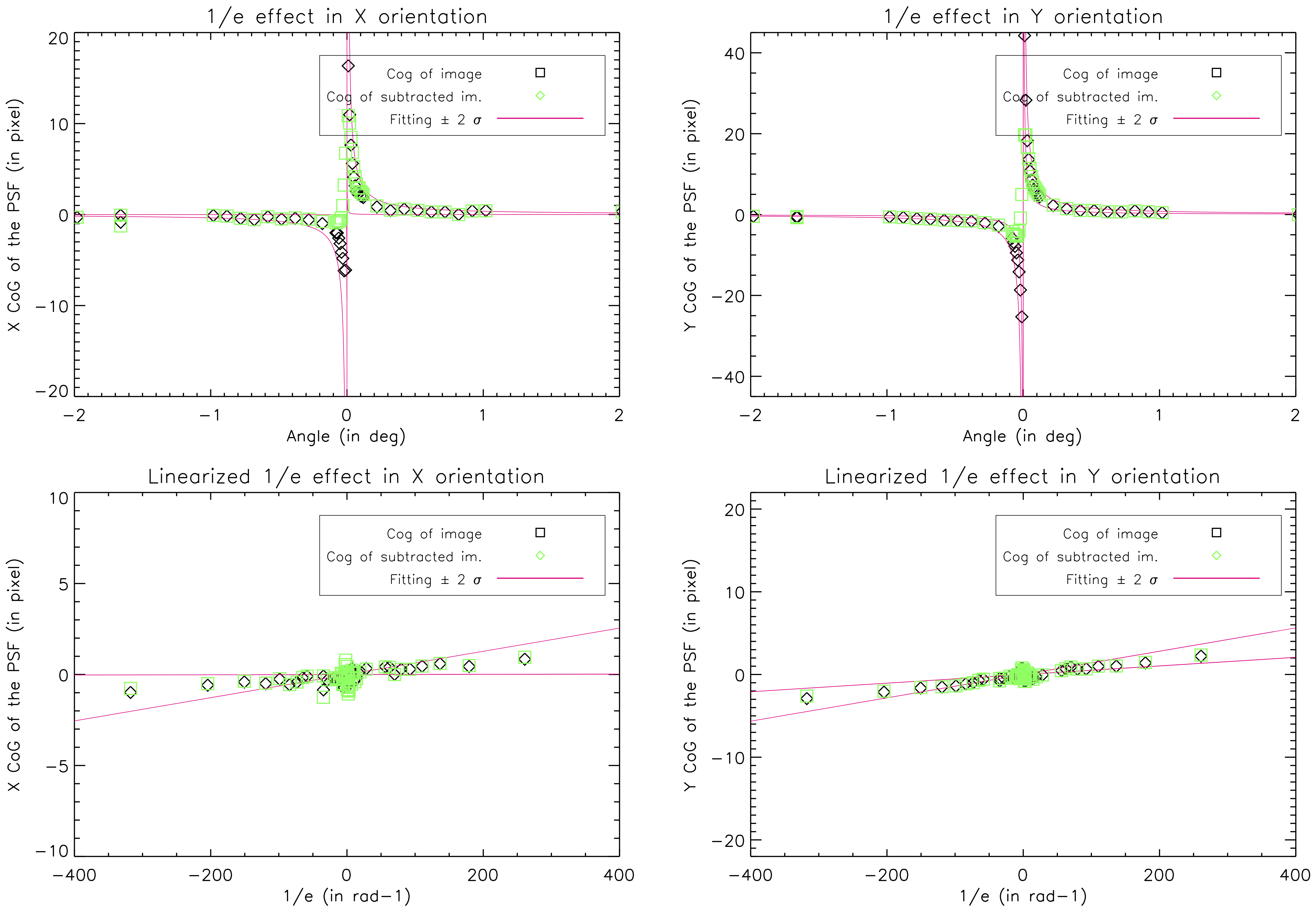}
    \caption{CoG in X and Y as a function of angle $\omega$ around crossed polarization angle $\omega_{\perp}$ for the case of a single mirror tested outside of the THD2 bench (corresponding to the 2 bottom rows in Fig. \ref{fig:ImagesGHIF_measurements}). Linear figures and inversed X-axis figures showing the fitting values with a $\pm 2 \sigma$ dispersion.}
    \label{fig:Weak_measurmt_single_mirror}
\end{figure}

\begin{figure}
\centering
    \includegraphics[width=0.9\linewidth]{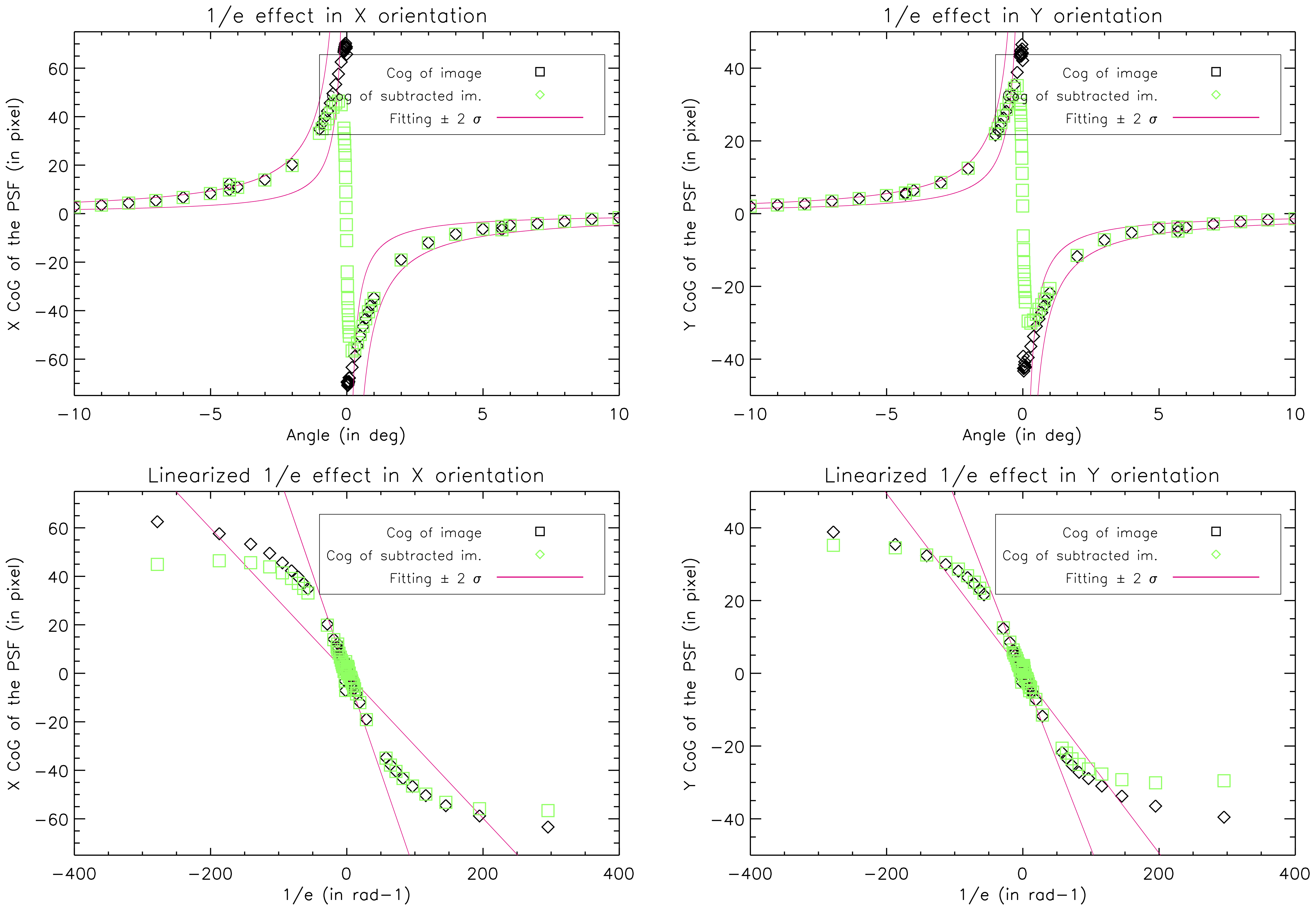}
    \caption{CoG in X and Y as a function of angle  $\omega$ around crossed polarization angle $\omega_{\perp}$ measured between FM2 and DM2. The data used is the same as the one shown on the 2D plot in Fig. \ref{fig:2D_cog} and in the 2 top rows of Fig. \ref{fig:ImagesGHIF_measurements}. Linear figures and inversed X-axis figures showing the fitting values with a $\pm 2 \sigma$ dispersion.}
    \label{fig:Weak_measurmt_FM2toDM2}
\end{figure}

The fitted positions are removed for each angle values and this processed  data is used to fit the amplification function given in Eq. \ref{eq:fit_g}. We fit this function by calculating a linear fit on the inverse function as shown in Fig. \ref{fig:Weak_measurmt_single_mirror} and Fig. \ref{fig:Weak_measurmt_FM2toDM2}.
This second fit is also calculated only on the angles that carry the right information. For the largest defects (like in Fig. \ref{fig:Weak_measurmt_FM2toDM2}), the fit is done for angles $\omega-\omega_{\perp}$ that are larger than $\pm 0.5\degree$ and smaller than $\pm 15\degree$ while we only use the angle larger than $\pm 0.1\degree$ and smaller than $\pm 1\degree$ for the smallest defects (like in Fig. \ref{fig:Weak_measurmt_single_mirror}).

\begin{table}[h!]
\resizebox{\textwidth}{!}{
   \centering 
 \begin{tabular}{l|c|c|c|c|c|c|c|c|c|c|c|c|c|c|c|c} 
        \hline \hline
         &\multicolumn{2}{|c}{Fiber to coro}&\multicolumn{2}{|c}{FM2 to Lyot}&\multicolumn{2}{|c}{FM2 to DM2}&\multicolumn{2}{|c}{DM2} & \multicolumn{2}{|c}{coro to Lyot}&\multicolumn{4}{|c}{FM2}&\multicolumn{2}{|c}{no mirror} \\
        \hline\hline
         $\lambda$ in nm &\multicolumn{2}{|c}{640, 705, or 785} & \multicolumn{2}{|c}{785} & \multicolumn{2}{|c}{640} & \multicolumn{2}{|c}{785} & \multicolumn{2}{|c}{640} & \multicolumn{2}{|c}{640} & \multicolumn{2}{|c}{785} & \multicolumn{2}{|c}{640} \\
\hline
            axis & H & V & H & V & H & V & H & V & H & V & H & V& H & V & H & V  \\
              
            mvt in nm& 623 & 227 & 540 & 290 & 551 & 240 & 555 & 340 & 2 & 5 & 4 & 32 & 3 & 60 & 20 & 24\\
            1$\sigma$ RMS & 70 & 70 & 70 & 45 & 70 & 75 & 130 & 70 & 20 & 20 & 4 & 5 & 5 & 10 & 3 & 3\\
        \hline
   
  \end{tabular}}
      \smallskip
    \caption{Recorded shift between horizontal and vertical polarization state inputs using the weak measurement method described in the text. The measurements have been done at different wavelengths and for different optics or group of optics on the THD2 bench. The signs of the value have been purposely removed because we did not keep track of the rotation angle of the analyzer with respect to the beam orientation.} \label{tab:GH&IF_Measured_values_THD_2}
\end{table}

The measured total shift induced from the fiber to the coronagraph is of the same order of magnitude ($\sqrt{623^2+227^2}=$663nm) as the shift measured in the DH images on the THD2 bench (800 nm, see \ref{sec:intro}). These values are much larger than the simulated effects shown in Table \ref{tab:GH&IF_predicted_values_THD_2}.
The FM2 shift effect is measured between 32 and 60 in the vertical direction, which is much smaller than the measured shift that is mostly seen in the horizontal direction.
The different shift measurements shown in Table \ref{tab:GH&IF_Measured_values_THD_2} indicates that the main contributor located between FM2 and the Lyot plane is DM2. Indeed, there is a clear difference between the measurements that includes DM2 ($>500$ nm in horizontal and $>200$ nm in vertical) and the other ones that does not include it (FM2 alone or corono to Lyot). Note that the measurements errors can be large (70  to 130 nm RMS) for the largest shifts. These large errors might explain, for example, why the vertical shift of DM2 alone is 100 nm larger than the measurement between FM2 and DM2. This difference could also be explain by a compensation of the DM2 shift by an another mirror.

This strong shift of the DM2 is very unexpected since DM2 has a simple aluminum coating and a much smaller incident angle than FM2 or DM1. 
The shift measured between the entrance fiber and the coronagraph is about 80 nm larger in horizontal than the measurements of the components measured after or including FM2 (DM2 alone, FM2 to DM2, FM2 to Lyot). The difference can probably be introduced by the entrance mirrors, especially DM1.
However, DM1 being very similar to DM2 and with a larger incident angle, it should also introduce a strong shift. So, either DM1 and DM2 have a different behavior or the effects do not add up linearly.
A specific measurement of the effect around DM1 is required to conclude but we have not been able to set it up because the space to fit a polarizer and an analyzer around DM1 is rather tight.

\begin{figure}[h!]
\centering
    \includegraphics[width=0.3\linewidth]{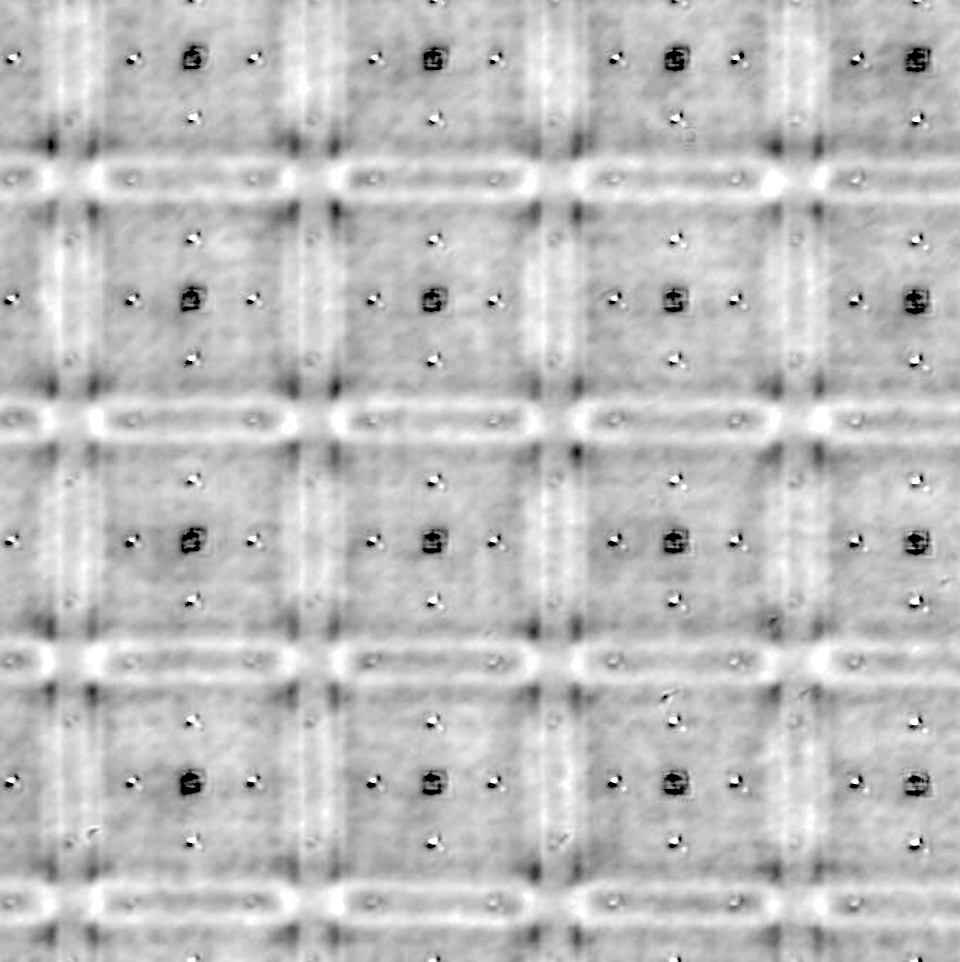}
    \caption{A 0.6 mm x 0.6 mm image of DM2 mirror measured using a dedicated imaging interferometer\cite{Mazoyer2014} showing high frequency structures.}
    \label{fig:DM_surface}
\end{figure}

The origin of the these strong polarization shift effects for DM2 is unclear. There is a $SiO_2$ window in front of the DM. Even, if it should not have any impact, we used our methodology of weak measurement to test one spare of this window between polarizer and an analyzer and found no clear effects above the typical noise estimated with a measurement without mirror. Since the level is much larger than what is expected from a metallic effect with a small angle ($6.5 \degree$):
\begin{itemize}
    \item Either the coating delivered by Boston Micromachine is not the one expected and it is made of dielectric layers that are known to create strong GH and IF effects.
    \item Or the non-symmetrical high frequency structures observed at micrometer level on the surface of the mirror (see Fig. \ref{fig:DM_surface}) have sub-wavelength impacts that introduce polarization effects, similar to metasurfaces.
\end{itemize}

\section{Summary and conclusion}
We showed that a Boston Micromachine deformable mirror located in the pupil on our bench (DM2) introduces a differential polarization shift between the PSF related to vertical and horizontal states. This shift has a strong impact on contrast for coronagraphs that are highly sensitive to tip-tilt effects like the FQPM and it does not allow reaching the best contrast with both polarization states simultaneously. 
The order of magnitude of this shift was estimated using coronagraphic images through the whole bench and over a series of measurements using cross-polarizers to amplify the small movement through weak measurement as described in Goswami 2014.\cite{Goswami}
The effect is not fully understood since the theoretical predictions are more than ten times smaller than what was found experimentally. Moreover, the effect should be limited by other mirrors than DM2. 
We need to verify the impact of the other DM that is on the bench (DM-1, Fig. \ref{fig:THD2_design}). In the case we find the same behavior, this will have an impact on current and future instrument designs. Indeed, while coronagraphs for current and future telescopes like Habitable World Observatory are much less sensitive to tip-tilt effects, this polarization shift, if it comes from the print-trough high frequency structures, will probably also introduce second-order polarization effects that could strongly limit the performance of these ambitious future instruments.
\acknowledgments 
 
 This work has benefited from the support of the French Action Sp\'ecifique Haute R\'esolution Angulaire (ASHRA) of CNRS/INSU co-funded by CNES.

\bibliography{spie2024} 
\bibliographystyle{spiebib} 

\end{document}